\documentclass[letterpaper]{jpconf}
\usepackage{graphicx}
\begin{document}
\title{Symmetry and Supersymmetry in Nuclear Pairing: Exact Solutions}

\author{A.B. Balantekin$^1$, Y. Pehlivan$^2$}

\address{$^1$ Physics Department, University of Wisconsin, Madison, WI  53706 USA}
\address{$^2$ Mimar Sinan Fine Arts University, Besiktas, Istanbul 34349, Turkey}

\ead{baha@physics.wisc.edu,yamac@physics.wisc.edu}

\begin{abstract}
Pairing plays a crucial role in nuclear spectra and attempts to describe it has a long history in nuclear physics. The limiting case in which all single particle  states are degenerate, but with different s-wave pairing strengths  was only recently solved. In this strong coupling limit the nuclear pairing Hamiltonian also exhibits a supersymmetry. Another solution away from those limits, namely two non-degenerate single particle states with different pairing strengths, was also given. In this contribution these developments are summarized and difficulties with possible generalizations to  more single particle states and d-wave pairing are discussed.
\end{abstract}

\section{Introduction}
Pairing plays a very prominent role in nuclear structure physics. The s-wave pairing problem can be
conveniently formulated using the quasi-spin operators:
\begin{equation}
\hat{S}^+_j=\sum_{m>0} (-1)^{(j-m)} a^\dagger_{j\>m}a^\dagger_{j\>-m},
\>\>\>\> \hat{S}^-_j=\sum_{m>0} (-1)^{(j-m)} a_{j\>-m}a_{j\>m},
\end{equation}
and
\begin{equation}
\hat{S}^0_j=\frac{1}{2}\sum_{m>0}
\left(a^\dagger_{j\>m}a_{j\>m}+a^\dagger_{j\>-m}a_{j\>-m}-1
\right) = \hat{N}_{j}-\frac{1}{2}\Omega_j,
\end{equation}
where $\Omega_j=j+\frac{1}{2}$ is the maximum number of pairs that can occupy the level $j$
and the number operator is
\begin{equation}
\hat{N}_j=\frac{1}{2}\sum_{m>0}
\left(a^\dagger_{j\>m}a_{j\>m}+a^\dagger_{j\>-m}a_{j\>-m}\right).
\end{equation}
Quasi-spin operators generate mutually commuting SU(2) algebras:
\begin{equation}
[\hat{S}^+_i, \hat{S}^-_j ] = 2 \delta_{ij} \hat{S}^0_j,
\>\>\>\>\>\>\> [\hat{S}^0_i, \hat{S}^{\pm}_j] = \pm \delta_{ij}
\hat{S}^{\pm}_j
\end{equation}
which are realized in the $s_j = \frac{1}{2}\Omega_j$ representations.  As we describe later, it is possible to define similar operators for d-wave pairing situations.

\section{Exact Solutions}

Over the years considerable attention was paid to exactly solvable pairing Hamiltonians with one- and two-body interactions.  These cases include
\begin{itemize}
\item The exact quasi-spin limit \cite{kerman1}:
\begin{equation}
\label{5}
\hat{H}=- |G|\sum_{jj'}  \hat{S}^+_j \hat{S}^-_{j'}.
\end{equation}
\item Richardson's solution for the case when the single particle energies are added to the
Hamiltonian in Eq.(\ref{5}) \cite{rich}
\begin{equation}
\label{6}
\hat{H}=\sum_{jm} \epsilon_j a^\dagger_{j\>m} a_{j\>m} -
|G|\sum_{jj'} \hat{S}^+_j \hat{S}^-_{j'}.
\end{equation}
\item A model of Gaudin's \cite{gaudin}, which is closely related to the Richardson's solution.
\item The limit with separable pairing in which the energy levels are degenerate
(the one-body term becomes a constant for a given number of pairs)
\cite{Pan:1997rw,Balantekin:2007vs,Balantekin:2007qr}:
\begin{equation}
\label{7}
\hat{H}=- |G|\sum_{jj'}c^*_jc_{j'} \hat{S}^+_j \hat{S}^-_{j'}.
\end{equation}
\item Most general separable case with two orbitals \cite{Balantekin:2007ip}.
\end{itemize}

Let us first examine the degenerate case. It is convenient to define the operators
\begin{equation}
\label{8}
\hat{S}^+(0)=\sum_j c^*_j\hat{S}^+_j \ \ \ \ \mbox{and} \ \ \ \ \
\hat{S}^-(0)=\sum_j c_j\hat{S}^-_j.
\end{equation}
Then the Hamiltonian in Eq. (\ref{7}) can be written as
\begin{equation}
\hat{H}=-|G|\hat{S}^+(0)\hat{S}^-(0).
\end{equation}
In the 1970's Talmi \cite{talmi} showed that
under certain assumptions, a state of the form
\begin{equation}
\label{10}
\hat{S}^+(0)|0\rangle=\sum_j c^*_j\hat{S}^+_j |0\rangle
\end{equation}
where $|0\rangle$ is the particle vacuum,
is an eigenstate of a class of Hamiltonians including the one above. Indeed
\begin{equation}
\hat{H}\hat{S}^+(0)|0\rangle= \left( -|G|\sum_j \Omega_j |c_j|^2
\right) \hat{S}^+(0)|0\rangle .
\end{equation}
How can we identify other one-pair states, which are eigenstates of the Hamiltonian in Eq.
(\ref{7})? For example for two levels $j_1$ and $j_2$, the state orthogonal to the one in Eq.
(\ref{10})
\begin{equation}
\left( \frac{c_{j_2}}{\Omega_{j_1}} \hat{S}^+_{j_1} -
\frac{c_{j_1}}{\Omega_{j_2}} \hat{S}^+_{j_2} \right) |0 \rangle,
\end{equation}
is also an eigenstate with E=0. It turns out that there is a systematic way to find such states. First
one introduces the operators
\begin{equation}
\label{13}
\hat{S}^+(x)=\sum_j\frac{c^*_j}{1-|c_j|^2x}\hat{S}^+_j, \>\>\>\>
\hat{S}^-(x)=\sum_j\frac{c_j}{1-|c_j|^2x}\hat{S}^-_j.
\end{equation}
(Note that the operators in Eq. (\ref{8}) are the same as the operators in Eq. (\ref{13}) calculated at $x=0$).  Then one can show that \cite{Pan:1997rw,Balantekin:2007vs}
\begin{equation}
\label{14}
\hat{S}^+(0)\hat{S}^+(z^{(N)}_1) \dots
\hat{S}^+(z^{(N)}_{N-1})|0\rangle
\end{equation}
is an eigenstate of the Hamiltonian in Eq. (\ref{7}) with energy
\begin{equation}
\label{15}
E_N =-|G|\left(\sum_j \Omega_j |c_j|^2-\sum_{k=1}^{N-1}
\frac{2}{z^{(N)}_k}\right)
\end{equation}
if the following Bethe ansatz equations
are satisfied:
\begin{equation}
\label{16}
\sum_j \frac{-\Omega_j/2}{1/|c_j|^2-z^{(N)}_m}
=\frac{1}{z^{(N)}_m}+\sum_{k=1(k\neq m)}^{N-1}
\frac{1}{z^{(N)}_m-z^{(N)}_k} \ \ \ \ \ \ \
m=1,2,\dots N-1.
\end{equation}
Similarly
\begin{equation}
\label{17}
\hat{S}^+(x^{(N)}_1)\hat{S}^+(x^{(N)}_2) \dots
\hat{S}^+(x^{(N)}_N)|0\rangle
\end{equation}
is an eigenstate with zero energy if the following Bethe ansatz equations
are satisfied:
\begin{equation}
\sum_j \frac{-\Omega_j/2}{1/|c_j|^2-x^{(N)}_m}=\sum_{k=1(k\neq
m)}^N \frac{1}{x^{(N)}_m-x^{(N)}_k} \ \ \ \ \ \ \mbox{for every} \
\ \ m=1,2,\dots,N .
\end{equation}
It should be emphasized that the states in Eqs. (\ref{14}) and
(\ref{17}) are eigenstates of the Hamiltonian in Eq. (\ref{7}) if
available single-particle levels are at most half full. One can show
that, if the single-particle levels are more than half full, the
state
\begin{equation}
\label{19}
\hat{S}^-(z_1^{(N)})\hat{S}^-(z_2^{(N)})\dots\hat{S}^-(z_{N-1}^{(N)})
|\bar{0}\rangle
\end{equation}
is an eigenstate with the same energy as in Eq. (\ref{15}) if the Bethe ansatz equations given in Eq.(\ref{16}) are satisfied \cite{Balantekin:2007vs}.  In Eq. (\ref{19})  $|\bar{0}\rangle$ designates the  state where all
single-particle levels are completely filled.

One should emphasize that although the solutions presented here are exact solutions, to obtain
explicit expressions for the energies and eigenstates, one still needs to obtain solutions of the Bethe ansatz equations. So far methods to solve the Bethe ansatz equations were developed  only in a limited number of cases  \cite{Balantekin:2007vs,Balantekin:2008ah}.

Note that it is also possible to calculate the quantum invariants
of these pairing Hamiltonians \cite{yamac}.

\section{Supersymmetry}

The above discussion indicates that the states of the Hamiltonian in Eq. (\ref{7}) with $N$ pairs
(Eq. \ref{14}) and with  $N_{max}+1-N $ pairs (Eq. \ref{17}) have the same eigenenergy. Furthermore the zero energy states are missing if the single-particle states are more than half full. This situation is reminiscent of the spectral conditions of the supersymmetric quantum mechanics \cite{Witten:1981nf}. Indeed introducing the  operator \cite{Balantekin:2007qr}
\begin{equation}
\hat{T} = \exp \left( - i \frac{\pi}{2} \sum_i (\hat{S}_i^+ + \hat{S}_i^-)
\right) ,
\end{equation}
which transforms the empty single-particle state, $|0\rangle$, to the fully
occupied state,
$|\bar{0}\rangle$:
\begin{equation}
\hat{T} |0\rangle=|\bar{0}\rangle ,
\end{equation}
one can define new operators
\begin{equation}
\hat{B}^- = \hat{T}^{\dagger} \hat{S}^-(0),\ \ \ \ \ \ \ \
\hat{B}^+ = \hat{S}^+(0) \hat{T}.
\end{equation}
Supersymmetric quantum mechanics tells us that the partner
Hamiltonians $\hat{H}_1 = \hat{B}^+ \hat{B}^-$ and $\hat{H}_2 =
\hat{B}^- \hat{B}^+$ have identical spectra except for the extremal
(usually ground) state of $\hat{H}_1$. Here two Hamiltonians
$\hat{H}_1$ and $\hat{H}_2$ are actually identical and equal to the
pairing Hamiltonian, Eq. (\ref{7}). Hence the role of the
supersymmetry is to connect the states in Eqs. (\ref{14}) and
(\ref{19}), i.e. this supersymmetry connects particle and hole
states.

\section{Exact solution for two single-particle states}

It turns out that one can find an exact solution for the case where there are only two single-particle
levels \cite{Balantekin:2007ip}:
\begin{equation}
\label{23}
\frac{\hat{H}}{|G|}= \sum_{j} 2\varepsilon_j
\hat{S}_j^0 - \sum_{jj'}c^*_jc_{j'} \hat{S}^+_j
\hat{S}^-_{j'}+\sum_j\varepsilon_j\Omega_j,
\end{equation}
where $\varepsilon_j$ and $c_j$'s are dimensionless and the sums are performed over only two
single-particle states.

The eigenstates of the Hamiltonian in Eq. (\ref{23}) can be written using the step operators:
\begin{equation}
{\cal J}^+(x)=\sum_j\frac{c_j^*}{2\varepsilon_j-|c_j|^2x}S_j^+
\end{equation}
as
\begin{equation}
{\cal J}^+(x_1){\cal J}^+(x_2)\dots {\cal J}^+(x_N)|0\rangle .
\end{equation}
Defining the auxiliary quantities
\begin{equation}
\beta=2\frac{\varepsilon_{j_1}-\varepsilon_{j_2}}{|c_{j_1}|^2-|c_{j_2}|^2}
\quad \quad \quad  \delta
=2\frac{\varepsilon_{j_2}|c_{j_1}|^2-\varepsilon_{j_1}|c_{j_2}|^2}
{|c_{j_1}|^2-|c_{j_2}|^2},
\end{equation}
one obtains the energy eigenvalues as
\begin{equation}
E_N=-\sum_{n=1}^N\frac{\delta x_n}{\beta-x_n}.
\end{equation}
In the above equations, the parameters $x_k$ are to be found by
solving the Bethe ansatz equations
\begin{equation}
\sum_{j}\frac{\Omega_j|c_j|^2}{2\varepsilon_j-|c_{j}|^2x_k}
=\frac{\beta}{\beta-x_k} +\sum_{n=1(\neq k)}^N\frac{2}{x_n-x_k} .
\end{equation}

\section{Prospects for other exact solutions}

We have shown that there is an exact solution of the s-wave pairing
problem with two non-degenerate orbitals in terms of the solutions
of the Bethe ansatz equations. (This problem is of course
numerically  diagonalizable in an SU(2) $\times$ SU(2) basis). It
seems to be very difficult to generalize the Bethe ansatz method to
the case of three non-degenerate orbitals. (Of course the three
single-particle level problem is diagonalizable in an SU(2) $\times$
SU(2) $\times$ SU(2) basis).

However, the Bethe ansatz method seems to be  generalizable to at least some d-wave pairing situations. In his work, mentioned earlier Talmi showed that, for a Hamiltonian with only one- and
two-body interactions, if the doubly-magic ground state energy is normalized to zero,
\begin{equation}
H |0 \rangle = 0,
\end{equation}
then the conditions
\begin{equation}
[H, S^{\dagger} (0) ] = V S^{\dagger} (0)
\end{equation}
and
\begin{equation}
[ [ H, S^{\dagger} (0) ], S^{\dagger} (0) ] = W \left( S^{\dagger} (0) \right)^2
\end{equation}
are sufficient to determine the entire energy spectra. In addition, if one considers d-wave pairing
operator
\begin{equation}
D^{\dagger}_M = \sum_{jj'} \alpha_{jj'} \left( a^{\dagger}_j \times a^{\dagger}_{j'} \right)^{(2)}_M,
\end{equation}
then the additional condition
\begin{equation}
[ [ H, S^{\dagger} ], D^{\dagger}_M ] = W  S^{\dagger} D^{\dagger}_M
\end{equation}
makes eigenstates with d-wave pairs possible
\begin{equation}
H D^{\dagger}_M |0 \rangle \propto D^{\dagger}_M |0 \rangle .
\end{equation}
Note that these conditions suggest existence of an algebraic structure. In fact, Ginocchio model
\cite{Ginocchio:1978tc} satisfies the double commutators given above.

One should explore if the Bethe ansatz method can be generalized in other cases where we also have an algebraic framework.
A possible answer lies in Gaudin's method mentioned earlier. It can be shown that the infinite-dimensional algebra
\begin{equation}
\label{35}
[J^+(\lambda),J^-(\mu)]=2\frac{J^0(\lambda)-J^0(\mu)}{\lambda-\mu},
\end{equation}
\begin{equation}
[J^0(\lambda),J^{\pm}(\mu)]=\pm\frac{J^{\pm}(\lambda)-J^{\pm}(\mu)}
                                                  {\lambda-\mu},
\end{equation}
\begin{equation}
[J^0(\lambda),J^0(\mu)]=[J^{\pm}(\lambda),J^{\pm}(\mu)]=0
\end{equation}
can be used to find eigenvalues and the eigenvectors of the Hamiltonian
\begin{equation}
\label{38}
H(\lambda)=J^0(\lambda)J^0(\lambda)+\frac{1}{2}J^+(\lambda)J^-(\lambda)+
           \frac{1}{2}J^-(\lambda)J^+(\lambda) .
\end{equation}
For details the reader is referred to, e.g. Refs. \cite{Balantekin:2005sj}, \cite{Balantekin:2005ks}, and
\cite{Balantekin:2007er}. (It should be emphasized that the Hamiltonian in Eq. (\ref{38}) is {\em not} the Casimir operator of the infinite-dimensional algebra given above).
The connection between the Hamiltonian in Eq. (\ref{38}) and the Hamiltonian
originally solved by Richardson (Eq. \ref{6}) is uncovered by using the realization
\begin{equation}
\label{39}
J^{0}(\lambda)=\sum_{i=1}^N\frac{\hat{S}^0_i}{\epsilon_i-\lambda}
\quad\mbox{and}\quad
J^{\pm}(\lambda)=\sum_{i=1}^N \frac{\hat{S}^{\pm}_i}{\epsilon_i-\lambda} ,
\end{equation}
where $S^{\pm}_i$ and $S^0_i$ are the quasi-spin operators. A possible procedure to include d-wave
pairing would first identify the underlying algebraic structure (i.e. the SO(8) symmetry of the Ginocchio Model), then introduce the analogs of the Eqs. (\ref{35}) through (\ref{38}) using a realization analogous to that in Eq. (\ref{39}). In fact a boson pairing model,
using the SU(1,1) analog of the quasi-spin in this fashion, was already studied in some detail
\cite{Balantekin:2005ks,Balantekin:2004yf}. However, such a program with arbitrary fermion pairings
remains to be largely unexplored.

One should finally remark that some of the techniques discussed here are more generally applicable
to other Hamiltonians with special one- and two-body interactions. One such example is the
Hamiltonian describing a dense electron gas near the center of a core-collapse supernova \cite{Balantekin:2006tg}.

\ack
This work was supported in part
by the U.S. National Science Foundation Grant No. PHY-0855082
and
in part by the University of Wisconsin Research Committee with funds
granted by the Wisconsin Alumni Research Foundation.

\end{document}